\documentclass[journal]{IEEEtran}

\usepackage[utf8]{inputenc}
\usepackage{cite}
\usepackage{amsmath,amssymb,amsfonts}
\usepackage{mathtools}
\usepackage{array}
\usepackage{algorithmic}
\usepackage{enumitem}
\usepackage{graphicx}
\usepackage{textcomp}
\usepackage{xcolor}
\usepackage{tikz}
\usepackage[american,RPvoltages]{circuitikz}
\usepackage{booktabs}
\usepackage[font=small]{caption}
\usepackage{subcaption}
\usepackage{wrapfig,booktabs,tabularx} %
\usepackage{multirow}
\usepackage{ltablex}
\usepackage{changepage}
\usepackage{floatrow}
\usepackage{placeins}
\usepackage{listings}
\usepackage[bookmarks]{hyperref}

\usepackage{adjustbox}

\newcommand\copyrighttext{%
  \footnotesize \textcopyright 2021 IEEE. Personal use of this material is permitted.
  Permission from IEEE must be obtained for all other uses, in any current or future
  media, including reprinting/republishing this material for advertising or promotional
  purposes, creating new collective works, for resale or redistribution to servers or
  lists, or reuse of any copyrighted component of this work in other works.}
\newcommand\copyrightnotice{%
\begin{tikzpicture}[remember picture,overlay]
\node[anchor=south,yshift=8pt] at (current page.south) {\fbox{\parbox{\dimexpr\textwidth-\fboxsep-\fboxrule\relax}{\copyrighttext}}};
\end{tikzpicture}%
}
\allowdisplaybreaks

\definecolor{color01}{rgb}{0.65098039,0.80784314,0.89019608}
\definecolor{color02}{rgb}{0.12156863,0.47058824,0.70588235}
\definecolor{color03}{rgb}{0.69803922,0.8745098,0.54117647}
\definecolor{color04}{rgb}{0.2,0.62745098,0.17254902}
\definecolor{color05}{rgb}{0.98431373,0.60392157,0.6}
\definecolor{color06}{rgb}{0.89019608,0.10196078,0.10980392}
\definecolor{color07}{rgb}{0.99215686,0.74901961,0.43529412}
\definecolor{color08}{rgb}{1,0.49803922,0}
\definecolor{color09}{rgb}{0.79215686,0.69803922,0.83921569}
\definecolor{color10}{rgb}{0.41568627,0.23921569,0.60392157}
\definecolor{color11}{rgb}{1,0.8,0}

\newfloatcommand{capbtabbox}{table}[][\FBwidth]
\usepackage{ragged2e}
\newcolumntype{L}[1]{>{\RaggedRight\hspace{0pt}}p{#1}}
\newcolumntype{Q}[1]{>{\RaggedLeft\hspace{0pt}}p{#1}}
\newcolumntype{C}[1]{>{\Center\hspace{0pt}}p{#1}}
\newcolumntype{R}[2]{%
    >{\adjustbox{angle=#1,lap=\width-(#2)}\bgroup}%
    l%
    <{\egroup}%
}

\newcommand*\rot{\rotatebox{90}}

\makeatletter
\newcommand{\specialcell}[1]{\ifmeasuring@#1\else\omit$\displaystyle#1$\ignorespaces\fi}
\makeatother

\begin{document}

\title{Low Frequency AC Transmission Upgrades\\with Optimal Frequency Selection}

\author{David~Sehloff,~\IEEEmembership{Student Member,~IEEE,}
        and~Line~Roald,~\IEEEmembership{Member,~IEEE,}%
\thanks{This material is based upon work supported by the New York Power Authority (NYPA), the New York State Energy Research and Development Authority (NYSERDA), and the National Science Foundation Graduate Research Fellowship Program under Grant No. DGE-1747503.}
}

\maketitle%
\copyrightnotice%
\graphicspath{{./fig/}}%
\begin{abstract}
The advantages of operating selected transmission lines at frequencies other than the standard 50 or 60 Hz are numerous, encompassing increased power transfer capacity and better utilization of existing infrastructure. While high voltage DC (HVDC) is by far the most well-established example, there has been an emerging interest low frequency AC (LFAC) transmission in applications ranging from offshore wind to railway systems and mining. In this paper, we investigate the use of LFAC as a transmission upgrade and propose models and analysis methods to determine the optimal choice of frequency. The paper first presents an optimal power flow model with frequency as a variable, assuming modular multilevel converters for frequency conversion. Using this model, we analyze LFAC as an embedded upgrade in a transmission system using existing lines. We quantify the system-wide advantages from improved power flow control and frequency reduction and find that an LFAC upgrade achieves similar and sometimes better results compared with HVDC upgrades. Finally, we analyze the factors which determine the optimal frequency for these upgraded transmission lines, and we demonstrate the benefits of changing the frequency in response to different system topologies and operating conditions.
\end{abstract}%

\section{Introduction}
\IEEEPARstart{I}{n} the development of alternating current (AC) electric power systems, one of the first design challenges was to select a standard electrical frequency.
Converting power across different frequencies has never been a trivial task, and the absence of a standard frequency would have posed overwhelming challenges in large scale electrification. Since the 1890s, when standard frequencies of 50 and 60 Hz were chosen and large power systems grew with these standards at their core \cite{Stilwell1899}, the question of how to select frequency has not been revisited. However, recent advances in large-scale power electronics have enabled the practical use of other frequencies on individual transmission lines within a 50 or 60 Hz power system.
The most common example of multi-frequency systems is high voltage DC (HVDC) transmission embedded in an AC grid. Advantages of HVDC include increased power transfer capacity and control of power flow \cite{Reed2019,genyin2004,Chatzivasileiadis2013} and ability to transfer power between interconnections that are not synchronized \cite{makarov2017models}. These benefits have prompted existing and planned installations of HVDC in Europe \cite{entsoe2019, ultranet}, in the US \cite{pacificintertie,transbaycable,ERCOTdcties,ErcotproposedDcTie} and across the world \cite{pracca1996itaipu,kumar2003three,graham2005hvdc}.

In this paper, we consider another highly promising class of multi-frequency systems, namely low frequency AC (LFAC) transmission. Specifically, we are interested in quantifying the benefits of LFAC transmission upgrades, where one or several existing AC lines are converted to LFAC to increase transmission capacity and add power flow control. LFAC upgrades involve only transmission lines, excluding loads and generators except in certain unique cases where a generation source is designed to operate at a non-standard frequency due to power electronics or electric machine design \cite{Rosewater2018}.

LFAC is known to increase the transmission capacity of stability constrained lines \cite{Wang1996}, and existing protection system technology can be leveraged to enable multi-terminal systems (in contrast to multi-terminal HVDC, which is hindered by protection challenges \cite{BLOND2016}). An area of ongoing research is the dynamic performance of low frequency transmission. While further research is needed, several results show that lower frequencies can improve voltage stability \cite{Ngo2016PESGM,Ngo2016TD}. LFAC can also eliminate the need for series compensation, mitigating the sub-synchronous resonance introduced by series capacitors.

LFAC has a long history as a design choice in railway systems \cite{steimel2012,laury2018}, and a growing body of work demonstrates increased capacity and control for offshore wind \cite{Funaki2000, Ruddy2016} and land-based transmission \cite{Wang1996}. In addition, LFAC shows distinct advantages for power flow in multi-terminal configurations, along with other factors discussed in Section \ref{sec:lfac}.

Using multi-frequency power systems with power electronics for frequency conversion provides new opportunities to more flexibly select and adjust the LFAC frequency, and several recent works have investigated the impact of the choice of frequency.
The optimal frequency for maximal power transfer has been shown to depend on transmission line properties \cite{lfacmodeling}, and \cite{SorianoRangel2018} showed the connection between the frequency and converter size and cost. %
At a system level, \cite{MEERE2017321} studied the installation and operating costs of low frequency offshore wind collection networks, introducing an optimization over frequency to minimize the cost of an offshore wind system.
A step towards quantitative analysis of the system-wide benefits of LFAC operations was made in \cite{Nguyen2016}, which formulated a power flow problem for a multi-frequency network with power flow control across the interfaces.
Follow-up work included HVDC lines in the power flow model \cite{Nguyen2019} and formulated an AC optimal power flow (OPF) with LFAC lines considering frequency as a fixed parameter \cite{Nguyen2019opf} and as an optimization variable \cite{nguyen2019optimal}. The authors demonstrated the benefit of the frequency optimization for a multi-terminal LFAC wind collector network.

These existing approaches focus on the application of multi-frequency networks for offshore wind generation but exclude others, such as long distance overhead transmission.
Here, we specifically investigate LFAC as a transmission upgrade that can increase capacity on congested lines (particularly focusing on long, stability-constrained lines) and improve utilization of the overall grid by using the frequency converters for power flow control. Our goal is to evaluate the impact that such LFAC upgrades can have on overall system operational cost. In addition, we compare the benefits of LFAC with those achieved through HVDC technology and provide a quantitative analysis of the optimal operating frequencies under different operational conditions and topologies. In order to assess these benefits, we develop a new formulation and implementation of a frequency-dependent AC OPF, which includes improved converter modeling and a better representation of stability-constrained lines relative to previously proposed models \cite{nguyen2019optimal}. Specifically, our contributions are the following:

\noindent\emph{(1)} We formulate a new frequency-dependent AC OPF model where frequency is treated as an optimization variable in the LFAC parts of the network. %
The proposed multi-frequency OPF is similar to the model in \cite{nguyen2019optimal} but employs a different objective (cost minimization instead of loss minimization) and a different converter model. Another important difference is that our model uses the polar form of the AC power flow equations, which enable the straight-forward implementation of angle difference constraints to represent steady-state stability limits. The implementation of our model is released as an open source software package along with this paper.

\noindent\emph{(2)} Using the above model, we provide a framework to separately assess the benefits of (i) power flow control, (ii) lowering the frequency and (iii) a combination of both. We find that power flow control can improve utilization both on the LFAC line and the surrounding AC grid, while the primary benefit of frequency control on overhead lines is an increase in the transmission capacity for stability constrained lines.

\noindent\emph{(3)} We further discuss a method to establish a fair comparison of LFAC and HVDC upgrades and conclude that LFAC provides similar and sometimes better results than HVDC, depending on the HVDC configuration.

\noindent\emph{(4)} Finally, we analyze how the optimal choice of frequency varies for different topologies and operational scenarios. In our test case, the optimal frequencies typically lie in the range where the critical transmission corridors are loaded to their thermal capacity. At higher frequency, critical lines are constrained by their angle difference, while at lower frequencies the voltage drop becomes a limiting factor.

The paper is organized as follows: Section \ref{sec:lfac} provides a qualitative summary of LFAC benefits, while Section \ref{sec:power_flow} models variable frequency power flow, Section \ref{sec:opf_formulation} gives the OPF formulation and Section \ref{sec:implementation} describes the software implementation. Section \ref{sec:nordic} introduces the test case, and Sections \ref{sec:lfac_benefits} and \ref{sec:optimal_frequency} analyze advantages of multi-frequency systems. %
Finally, Section \ref{sec:conclusions} concludes. %

\section{Benefits and Challenges of LFAC Transmission}\label{sec:lfac}
The distinct advantages of LFAC transmission can outweigh the challenges and
motivate the rest of this work in analyzing multi-frequency systems. This section briefly describes these advantages and challenges.
\subsection{Benefits of LFAC Transmission}
\subsubsection{Increased Capacity on Stability Constrained Lines}
An important limit for long distance transmission is the transient stability limit \cite{kundur1994}. This is particularly important in systems with long lines, which may be prone to voltage collapse \cite{van2015test}. Under the lossless approximation, active power flow $p_{ij}$ between buses $i$ and $j$ with inductance $L_{ij}$ is ${p_{ij}=\frac{V_iV_j}{\omega L_{ij}}\sin\left(\theta_{ij}\right)}$ \cite{bergen2000}. The maximum power transfer is achieved when $|\theta_{ij}|=90^\circ$, though a maximum difference in voltage angle $< 90^\circ$ is typically imposed to ensure a margin for stable operation. However, decreasing the frequency $\omega$ reduces reactance $X_{ij}=\omega L_{ij}$, enabling greater power transfer, discussed in \cite{Wang1996,Funaki2000} and in more detail in \cite{Ngo2016PowSys,lfacmodeling}.

\subsubsection{Increased Capacity on Thermally Constrained Cables}
Cables and overhead lines are subject to thermal limits, represented as a maximum current or apparent power. In cables, charging currents from large shunt capacitive susceptance decrease the active power transmission allowable under the thermal limit. This susceptance $B_{ij}$ on a cable with capacitance $C_{ij}$ is $B_{ij}=\omega C_{ij}$. Lowering the frequency directly decreases the susceptance, thereby reducing the charging current. The resulting higher allowable power transfer \cite{Funaki2000} is a motivation for LFAC offshore wind power networks \cite{Ruddy2016}.

\subsubsection{Improved Power Flow Control}
Utilizing the frequency converter to control power flow across the interface offers another significant advantage, similar to the advantages demonstrated for Flexible AC Transmission (FACTS) and HVDC \cite{Chatzivasileiadis2011, Chatzivasileiadis2015}.
In the context of LFAC, the ability to control active power flow has been discussed \cite{Funaki2000,meliopoulos2012} and demonstrated \cite{Wang2006} for cycloconverters. Independent active and reactive power control has been modeled in steady state for back-to-back converters \cite{Nguyen2016} and simulated for back-to-back and matrix modular multilevel converters (MMC) \cite{CastilloSierra2020}. This control allows better utilization of both the LFAC and AC lines and thus extends the benefits of LFAC beyond increasing the capacity of a single branch.

\subsubsection{Utilization of Existing Infrastructure}
A benefit of LFAC is that the conductors already in place can be used with minimal or no upgrades.
Efficiency of existing equipment can also be improved, notably in cables, which experience lower sheath and dielectric losses at low frequencies \cite{Fischer2012}.

\subsubsection{Support of Multi-Terminal Systems}
A main advantage of LFAC relative to HVDC is the availability of protection systems, which enables the support of multi-terminal LFAC systems.
Though challenges include longer time between zero crossings and larger fault current due to smaller reactance, these can be overcome with extensions of existing AC protection technology \cite{Fischer2012}. In contrast, DC circuit breakers remain a challenge requiring further research.

\subsection{Challenges of LFAC Transmission}
With these benefits come several challenges. Frequency converters, which are expensive and large power electronics stations, are required at each interface between standard and low frequency lines. Connecting existing generation and loads directly to a low frequency line would require additional converters to interface the low frequency to the standard frequency of the generators and loads. If substations are present in a low frequency corridor, their components must be designed for the low frequency. Finally, transformers on low frequency branches are not desirable due to their large size, though if necessary they are technically feasible and commercially demonstrated \cite{Fischer2012}. These challenges and tradeoffs point to the need for planning tools which quantify the system-wide impacts of LFAC upgrades.

\section{Variable Frequency Power Flow Modeling}\label{sec:power_flow}
In this section, we model power flow in a network with multiple frequencies as variables. This is similar to the formulation in \cite{nguyen2019optimal}, but differences here include %
(i) the use of a polar coordinate AC power flow formulation with angle difference constraints on stability-constrained lines which enables analysis of the key advantages of LFAC, and (ii) modeling of modular multilevel converters (MMCs) which are advantageous for large-scale power conversion.
We use a similar lumped parameter $\Pi$ model but include the shunt conductance, as its effects become more significant with decreasing frequency.

\subsection{Multi-Frequency Network Modeling and Notation}
We define the parts of the network which share a frequency as \emph{subnetworks}, which exchange power through \emph{frequency conversion interfaces}. The set of all subnetworks is denoted by $\mathcal{S}$. Every subnetwork $l\in\mathcal{S}$ has the following properties:
	\subsubsection{Frequency} Subnetwork $l$ has a single frequency $\omega_l$.
    Sets $\mathcal{S}_\text{vf}$ and $\mathcal{S}_\text{cf}$ comprise subnetworks with variable and constant frequency, respectively. %
	\subsubsection{Components} The set of all buses in subnetwork $l$ is $\mathcal{N}_l$, and the set of all edges is $\mathcal{E}_l$. Each branch is represented as two directional edges, indexed as $ije$, where $i$ and $j$ are the origin and destination buses, respectively, and $e$ is the branch index. The set of edges from bus $i$ is $\mathcal{E}_{l,i}$. The set of all generators in subnetwork $l$ is $\mathcal{G}_{l}$ and of those connected to bus $i$ is $\mathcal{G}_{l,i}$.
	\subsubsection{Reference Bus} Subnetwork $l$ has a single reference bus which provides an angle reference for the subnetwork. The set of all reference buses in the network is $\mathcal{N}_\text{ref}$.
	\subsubsection{Interfaces} Two subnetworks can exchange power through a frequency conversion interface. The set of all frequency conversion interfaces is $\mathcal{I}$. Each interface $m\in\mathcal{I}$ has one converter, and the subscript $m$ indicates an individual converter. The set of buses directly connected to $m$ is $\mathcal{N}_m^I$.

\begin{figure}[!t]
	\centering
	\includegraphics[width=0.9\textwidth]{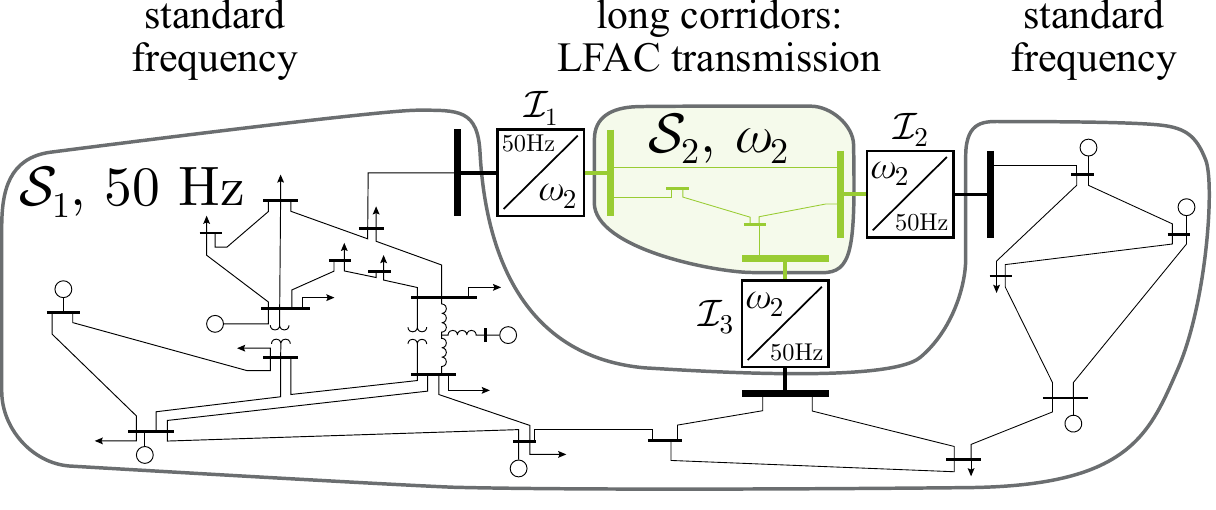}
	\caption{Network modeling framework for an example case with LFAC lines in a long corridor.}
	\label{fig:network}
\end{figure}

This framework allows any number of subnetworks and interfaces between them, as depicted in Fig. \ref{fig:network}. In this example, the main subnetwork, $\mathcal{S}_1$, has a standard fixed frequency of 50 Hz, and an LFAC corridor $\mathcal{S}_2$ has frequency $\omega_2$ considered to be a variable. We note that our modeling framework is able to handle more than two subnetworks connected through a single interface but restrict our discussion to interfaces connecting pairs of subnetworks.

\subsection{Frequency Conversion}

MMCs offer many advantages for LFAC, overcoming limitations of cycloconverters and two-level voltage source converters, including narrow ranges of achievable frequencies, lack of independent active and reactive power control, and filter and transformer needs \cite{Flourentzou2009,Allebrod2008,ludois2010}. They can be scaled to high voltage and power levels, demonstrated in practice for HVDC \cite{Westereweller2010} and AC-AC conversion for railways \cite{ABB2018}.

\begin{figure}[!t]
 \centering
 \resizebox{0.8\textwidth}{!}{%
\begin{circuitikz}[scale=0.8, american]
	\ctikzset{blocks/scale=1.6}
		\draw(0,0) node[fourport](mmc) {MMC}
		(mmc.port3) ++(0,0.5) coordinate(p3) to[generic, l^=$Z_j$] ++(4,0)
		(mmc.port2) ++(0,-0.5) coordinate(p2) to[short] ++(4,0)
		(p3) ++(4,0) coordinate(j) to[open, v^=, o-o] (p2 -| j)
		(p3) to[open, v_=$\underline{V}_j$, american open voltage=legacy] (p2 -| p3)
		(mmc.port4) ++(0,0.5) coordinate(p4) to[generic, l_=$Z_i$] ++(-4,0)
		(mmc.port1) ++(0,-0.5) coordinate(p1) to[short] ++(-4,0)
		(p4) ++(-4,0) coordinate(i) to[open, v_=, o-o] (p1 -| i)
		(p4) to[open, v^=$\underline{V}_i$, american open voltage=legacy] (p1 -| p4);
    \ctikzset{resistors/scale=0.5}
    \draw (p3) ++(1,0) coordinate(pp3) to[generic, l_=$\frac{Y^\mathrm{sh}_j}{2}$] (p2 -| pp3);
    \draw (p3) ++(3,0) coordinate(ppp3) to[generic, l_=$\frac{Y^\mathrm{sh}_j}{2}$] (p2 -| ppp3);
		\draw (p4) ++(-1,0) coordinate(pp4) to[generic, l=$\frac{Y^\mathrm{sh}_i}{2}$] (p1 -| pp4);
    \draw (p4) ++(-3,0) coordinate(ppp4) to[generic, l=$\frac{Y^\mathrm{sh}_i}{2}$] (p1 -| ppp4);
	\draw [draw=orange] (j -| i) ++(0.5,1.1) rectangle ($ (j) + (-0.5,-3.5) $);
	\node[above, text=orange] at ++(0,2.45) {converter $\mathcal{I}_m$};
	\draw (0,0.75) to node[nigbt, bodydiode, scale=0.8,rotate=90,color=black!30]{} (0,0.75);
	\draw (p4) ++(0,0.5) node[above] {$p_{im}, q_{im}$}
	(p4) ++(0,0.25) node[above] {$\longleftarrow$};
	\draw (p3) ++(0,0.5) node[above] {$p_{jm}, q_{jm}$}
	(p3) ++(0,0.25) node[above] {$\longrightarrow$};
\end{circuitikz}
}
	\caption{Single phase equivalent model of a frequency converter interface with optional filter or transformer branches.}%
	\label{fig:converter}
\end{figure}
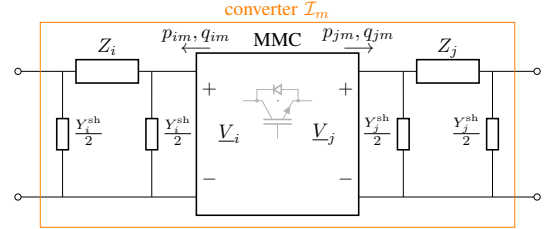

\begin{figure}[!t]
	\input{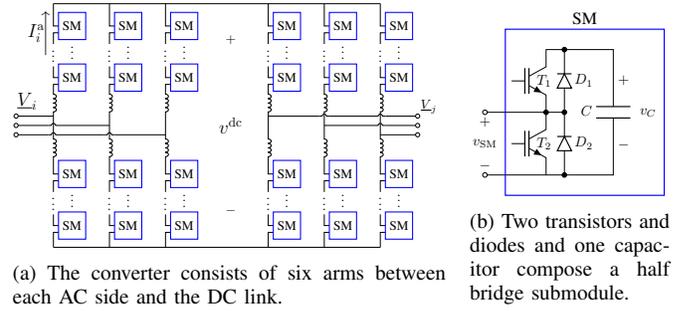}
	\caption{The back-to-back modular multilevel converter topology.} %
	\label{fig:mmc}
\end{figure}

For our analysis, we model the MMC as a two-port element as seen in Fig. \ref{fig:converter}, with $\Pi$ branches at the terminals to represent filters or transformers, if present.
Fig. \ref{fig:mmc} illustrates the more detailed topology of a back-to-back MMC, consisting of two AC-DC converter stages, each made up of six arms, or strings of semiconductor submodules in series.
Our model of the losses and limits is based on this topology and is explained below.

\subsubsection{Converter Losses}
The converter losses consist of conduction and switching losses and losses in the $\Pi$ branches.
\noindent\underline{Conduction losses:}
The semiconductor devices in each submodule are two diodes and two transistors, commonly IGBTs. The conduction loss in each of these devices depends on the on-state voltage drop, resistance, and current.
The total conduction loss is a function of the current through each arm $I^\mathrm{a}$ (specifically the squared rms $(I^\mathrm{a,rms})^2$ and the mean absolute value $I^\mathrm{a,mabs}$) along with the DC current, $I^\mathrm{dc}$ \cite{Yang2019}. For the bus $i$ terminal of converter $m$, these currents are%
\begin{align}
	&\textstyle(I^\mathrm{a,rms}_i)^2\!=\textstyle\frac{1}{18V_i^2}\left(M^2p_{im}^2+\frac{p_{im}^2+q_{im}^2}{2}\right), \label{eq:a_rms}
\\
	&\textstyle I^\mathrm{a,mabs}_i\!=\!\frac{2\sqrt{2}}{3 V_i}\!\left(\!\frac{M^2p_{im}^2}{\sqrt{p_{im}^2+q_{im}^2}}\!+\sqrt{\!q_{im}^2+p_{im}^2(\!1\!-\!M^2\!)}\!\right), \label{eq:a_mabs}
\\
	&\textstyle I^\mathrm{dc}_i\!=\textstyle\frac{M|p_{im}|}{\sqrt{2}V_i}. \label{eq:a_dc}
\end{align}
Here $p_{im}$ is the active power, $q_{im}$ the reactive power and $V_i$ the rms voltage magnitude. $M$ is the modulation index, which relates $V_i$ to the DC voltage $v^\mathrm{dc}$ as $M=\frac{\sqrt{2}V_i}{v^\mathrm{dc}}$.
The contributions of each of these currents to the conduction loss depend on the individual semiconductor parameters and can be represented by constant coefficients $C^{\textrm{c}1}$, $C^{\textrm{c}2}$, and $C^{\textrm{c}3}$. Ref. \cite{Yang2019} demonstrated accurate calculation of these coefficients from common datasheet parameters. The total conduction loss $p^\mathrm{cond}_{im}$ for the bus $i$ terminal of converter $m$ is the sum of the loss in all six arms and is given by
\begin{multline}
	p^\mathrm{cond}_{im}\! =\! 6\left(\!C^{\textrm{c}1}(I^\mathrm{a,rms}_i)^2\!+C^{\textrm{c}2}I^\mathrm{a,mabs}_i\!+C^{\textrm{c}3}I^\mathrm{dc}_i\!\right),\\\forall i\in \mathcal{N}_m^I, m\in\mathcal{I}\label{eq:cond_loss}.
\end{multline}

\noindent\underline{Switching losses:}
The energy lost in a turn-on or turn-off switching event depends on the conduction current of the semiconductor. The total switching losses in the converter, as an average power loss, depend on the squared rms current $(I^\mathrm{a,rms})^2$ and the mean absolute value of current $I^\mathrm{a,mabs}$ through each arm.
The contribution of each of these components to the total switching losses are expressed as constant coefficients $C^{\textrm{s}1}$, $C^{\textrm{s}2}$, and $C^{\textrm{s}3}$ and can be calculated using datasheet values \cite{Yang2019}. The total switching loss $p^\mathrm{sw}_{im}$ for the bus $i$ terminal of converter $m$ is given by
\begin{equation}
	p^\mathrm{sw}_{im}\!=\! 6\!\left(\!C^{\textrm{s}1}\!(\!I^\mathrm{a,rms}_i\!)^2\!+\!C^{\textrm{s}2}\!I^\mathrm{a,mabs}_i\!+\!C^{\textrm{s}3}\!\right),\!\forall i\!\in\!\mathcal{N}_m^I, m\!\in\!\mathcal{I}\label{eq:switch_loss}.
\end{equation}

\noindent\underline{$\Pi$ branch losses:}
Any filter elements or transformers can be represented by their $\Pi$ equivalent circuits. The parameters $Z_i$ and $Y_i$ represent either the filter inductance and capacitance \cite{Allebrod2008} or transformer magnetizing and leakage inductance, core loss and winding resistance.
Losses on all branches in the system model are included in the standard power flow equations, as discussed below.

\subsubsection{Converter Limits}
The maximum voltage of the MMC depends on the number of submodules in each arm and the voltage rating of each submodule. This imposes a maximum converter voltage, $\overline{V}$, on the voltage magnitude at each converter terminal bus:
\begin{align}
	V_i&\leq\overline{V_{i}}, \quad\forall i\in\mathcal{N}_m^I, m\in\mathcal{I}\label{eq:conv_voltage}.
\end{align}
Similarly, submodule ratings determine the maximum rms current in each arm $\overline{I^\mathrm{a,rms}}$:
\begin{equation}
	(I^\mathrm{a,rms}_i)^2-\overline{I^\mathrm{a,rms}_i}^2\leq0, \quad\forall i\in\mathcal{N}_m^I, m\in\mathcal{I}\label{eq:conv_current}.
\end{equation}
The power flow through the transformer or filter connected to either terminal of the converter is limited by %
standard branch thermal limits, as discussed below.

\subsubsection{Subnetwork Interface Power Balance}
For each interface, we enforce conservation of active power considering the power injected by the converter at both ports, $p_{im}^{I}$ and $p_{jm}^{I}$, and the conduction and switching losses from \eqref{eq:cond_loss} and \eqref{eq:switch_loss}:
\begin{equation}
	p_{im}^{I} \!+\! p_{jm}^{I} \!+\! p^\mathrm{cond}_{im} \!+\!p^\mathrm{cond}_{jm}+ p^\mathrm{sw}_{im}\!+\! p^\mathrm{sw}_{jm}\!=\!0,\quad\forall m \in \mathcal{I}.\label{eq:interface_pbal}
\end{equation}
The embedded energy storage in the MMC enables independent control of reactive power injection at each terminal. This reactive power is also independent of the active power, provided that the arm rms currents remain below their limits.

\subsubsection{Subnetwork Frequency}
If a subnetwork has variable frequency, $\omega_l$ must remain within the allowable range:
\begin{equation}
	\underline{\omega_l} \leq \omega_l \leq\overline{\omega_l},\qquad \forall l\in\mathcal{S}_\text{vf}\label{eq:var_f}.
\end{equation}
Here, $\underline{\omega_l}$ and $\overline{\omega_l}$ represent the minimum and maximum frequencies, respectively. These may be influenced by factors such as converter design and protection system limitations.

\subsection{Branch Parameters}
Next, we discuss the branch parameters for overhead lines. Note that we do not discuss LFAC cables in this paper. The lumped parameter $\Pi$ branch model was demonstrated and validated for nonstandard frequencies in \cite{lfacmodeling}.
In this model, branch $ije$ from bus $i$ to $j$ has lumped series resistance $R_{ije}$ which can be assumed constant on overhead lines over the range of considered frequencies.
The lumped series reactance $X_{ije}=\omega_l L_{ije}$ is a function of the lumped inductance $L_{ije}$ (independent of frequency) and frequency itself. Accordingly, the conductance $G_{ije}$ and susceptance $B_{ije}$ are both functions of the frequency:
\begin{multline}
		G_{ije} = \frac{R_{ije}}{R_{ije}^2+\omega_{l}^2 L_{ije}^2};\quad B_{ije} = -\frac{\omega_{l} L_{ije}}{R_{ije}^2+\omega_{l}^2 L_{ije}^2},\\\forall  ije\in\mathcal{E}_l, l\in\mathcal{S}.\label{eq:conductance}
\end{multline}
The shunt conductance is independent of frequency,
\begin{align}
	G_{ije}^\textrm{sh} &= \frac{1}{R_{ije}^\textrm{sh}}, &\qquad\forall  ije\in\mathcal{E}_l, l\in\mathcal{S}\label{eq:conductance_ch}
\end{align}
while shunt capacitor or reactor susceptance depends on the frequency and shunt capacitance or inductance, respectively,
\begin{equation}
	B_{ije}^\textrm{sh} = \omega_l C_{ije}^\textrm{sh}; \text{ or }B_{ije}^\textrm{sh} = \frac{-1}{\omega_l L_{ije}^\textrm{sh}},\quad\forall  ije\in\mathcal{E}_l, l\in\mathcal{S}.\label{eq:susceptance_ch}
\end{equation}
These equations also apply to shunt elements directly connected to buses, where the subscript denotes the bus.

We model each transformer as an ideal frequency-independent voltage step and phase shift connected to a $\Pi$ branch representing its internal characteristics. The off-nominal turns ratio is $\tau$ and the phase shift is $\phi$.

\subsection{Power Flow Equations}
The power flow on each branch is represented by the AC power flow equations with active and reactive power $p_{ije}$ and $q_{ije}$ injected from bus $i$:
\begin{align}
	\textstyle
  \MoveEqLeft[4] p_{ije}^{E}=\frac{V_i^2}{\tau_{ije}^2}\!\left(G_{ije}\! + \!G_{ije}^\textrm{sh}\right)
  \!-\!\frac{V_i V_j}{\tau_{ije}}\Big(G_{ije}\cos(\!\theta_i\!-\!\theta_j\!-\phi_{ije}\!)\notag\\&+B_{ije}\sin(\!\theta_i\!-\!\theta_j\!-\!\phi_{ije}\!)\Big),\forall ije\in\mathcal{E}_l, l\in\mathcal{S},\label{eq:p_o}\\
	\textstyle
	\MoveEqLeft[4] q_{ije}^{E}=-\frac{V_i^2}{\tau_{ije}^2} \left(B_{ije}+B_{ije}^\textrm{sh}\right)
	-\frac{V_i V_j}{\tau_{ije}}\Big(G_{ije}\sin(\!\theta_i\!-\theta_j\!-\phi_{ije})\notag\\&-B_{ije}\cos(\!\theta_i\!-\theta_j\!-\phi_{ije}\!)\Big),\forall ije\in\mathcal{E}_l, l\in\mathcal{S}\label{eq:q_o}.
\end{align}
To enforce nodal power balance, the active and reactive power entering each bus must each sum to zero:
\begin{align}
	\sum_{\mathclap{g\in \mathcal{G}_{l,i}}}\!p_g^{G}\!-\!\sum_{\mathclap{ije \in \mathcal{E}_{l,i}}}\!p_{ije}^{E}- p_i^L - V_i^2G_i^\text{sh}\!=\! 0, \quad\forall i\in\mathcal{N}_l, l\in\mathcal{S},\label{eq:pbal}\\
	\sum_{\mathclap{g\in \mathcal{G}_{l,i}}}\!q_g^{G} \!-\! \sum_{\mathclap{ije \in \mathcal{E}_{l,i}}}\!q_{ije}^{E} - q_i^L + V_i^2B_i^\text{sh} \!=\!0, \quad\forall i\in\mathcal{N}_l, l\in\mathcal{S}.\label{eq:qbal}
\end{align}
Here the active and reactive power load at each bus is denoted by $p_i^L$ and $q_i^L$, shunt conductance is $G_i^\text{sh}$, and shunt susceptance is $B_i^\text{sh}$.
The reference buses have zero voltage angles,
\begin{equation}
	\theta_i = 0, \qquad\forall i\in\mathcal{N}_\text{ref}\label{eq:vref}.
\end{equation}
In addition to the power flow equations, we include constraints to enforce limits on the branch flows, voltage magnitudes and generation capacity.

\noindent\underline{Stability limits:}
For transient stability, the absolute angle difference of connected buses must be less than $\overline{\theta}$:
\begin{equation}
	-\overline{\theta}\leq\theta_i - \theta_j\leq\overline{\theta}, \qquad\forall ije\in\mathcal{E}_l, l\in\mathcal{S}.\label{eq:ang_lim}
\end{equation}
\noindent\underline{Thermal limits:}
Thermal limits give the maximum apparent power, $\overline{s_{ije}^{E}}$, applied to power injections at each end of all lines:
\begin{align}
	\left(p_{ije}^{E}\right)^2 + (q_{ije}^{E})^2 \leq \left(\overline{s_{ije}^{E}}\right)^2,\qquad\forall ije\in\mathcal{E}_l, l\in\mathcal{S}.\label{eq:s_lim_f}
\end{align}
\noindent\underline{Bus voltage limits:}
The voltage magnitude at each bus must be between the lower and upper limits, $\underline{V_i}$ and $\overline{V_i}$:
\begin{equation}
	\underline{V_i}\leq V_i\leq\overline{V_i}, \qquad \forall i\in\mathcal{N}_l, l\in\mathcal{S}\label{eq:vlim}.
\end{equation}
\noindent\underline{Generator limits:}
Each generator has upper and lower limits for active and reactive power, $\overline{p_g^{G}}$, $\overline{q_g^{G}}$ and $\underline{p_g^{G}}$, $\underline{q_g^{G}}$:
\begin{equation}
	\underline{p_g^{G}}\leq p_g^{G}\leq\overline{p_g^{G}};\quad
	\underline{q_g^{G}}\leq q_g^{G}\leq\overline{q_g^{G}},\qquad\forall g\in\mathcal{G}_l, l\in\mathcal{S}.\label{eq:gen_pq}
\end{equation}

\section{Variable Frequency Optimal Power Flow}\label{sec:opf_formulation}%
We consider an economic dispatch problem \cite{ferc2005,Chowdhury1990} which focuses on operating the system at lowest generation cost. Each generator $g$ has a cost function $c_g(\cdot)$ which is a piecewise linear or polynomial function of the active power produced by the generator, and the objective is to minimize the total generator costs:
\begin{equation}
c_\mathrm{total} = \sum_{l\in \mathcal{S}}\sum_{g\in \mathcal{G}_l}c_g\left(p_g^{G}\right).\label{eq:obj}
\end{equation}
Combining this objective with the above variable frequency modeling, we obtain the full AC OPF problem with multiple, variable frequencies:

\begin{gather}
\begin{aligned}
	\multicolumn{2}{l}{$\displaystyle\min_{V,\theta,p^G,q^G,\omega,p^I,q^I}\quad c_\mathrm{total}$}&&(\ref{eq:obj})\\
	\mathrm{s.t.}\quad &\text{converter voltage and current limits}&&(\ref{eq:conv_voltage},\ref{eq:conv_current})\\
	&\text{converter power balance}&&(\ref{eq:interface_pbal})\\
	&\text{frequency constraints}&&(\ref{eq:var_f})\\
	&\text{active and reactive power flow}&&(\ref{eq:p_o},\ref{eq:q_o})\\
	&\text{branch angle and apparent power}&&(\ref{eq:ang_lim}, \ref{eq:s_lim_f})\\
	&\text{active and reactive power balance}&&(\ref{eq:pbal}, \ref{eq:qbal})\\
	&\text{bus voltage magnitude and angle ref}&&(\ref{eq:vlim}, \ref{eq:vref})\\
	&\text{generator active and reactive power}&&(\ref{eq:gen_pq})
\end{aligned} \label{eq:opf}\tag{$\star$}
\end{gather}
The formulation \eqref{eq:opf} is more complex than the standard AC OPF. It shares the standard decision variables corresponding to control of generator active power and voltage magnitude, but additional decision variables are introduced: the frequency of variable frequency lines and active and reactive power at the converter terminals are decision variables. The power flow on each LFAC branch has a cubic dependence on frequency, and the reactive power balance has a linear dependence on frequency at buses with shunts. Converters also add active and reactive power variables and nonlinear constraints for their losses and operating limits.

\section{Implementation}\label{sec:implementation}
The variable frequency AC OPF formulation is implemented in the Julia language \cite{bezanson2017julia} using the mathematical programming package JuMP \cite{DunningHuchetteLubin2017}.
We utilize the PowerModels package \cite{coffrin2018} to parse standard network data in the formats used by PSS\textregistered E or Matpower, which we combine with additional data including subnetworks, interfaces, frequency constraints and converter parameters.
The multi-frequency formulation requires the definition of series inductance and shunt capacitance, rather than reactance and susceptance as provided by standard network data. We therefore use the base frequency to calculate the inductance and capacitance and define frequency-dependent admittance values.

The data above is used to construct a JuMP model for the OPF formulation.
The resulting continuous, nonlinear, constrained optimization problem is solved using the interior point solver IPOPT \cite{wachter2006implementation}. To validate this implementation, we solve the standard fixed frequency AC OPF for each network in the OPF benchmarking library PGLib-OPF \cite{pglib}, which includes realistic networks with up to 30,000 buses. We compare the solutions with those found by PowerModels and find that the objective values agree to within $10^{-3}\%$ in all cases, and the variables agree similarly.

To enable other researchers to pursue LFAC studies and ensure reproducibility of our results, we are making the full code implementation available as the open source Julia package VariableFrequencyOPF.jl\cite{VariableFrequencyOPF}.

\section{The Nordic Test System}\label{sec:nordic}
\begin{figure}[!t]
	\centering
	\includegraphics[width=0.8\textwidth]{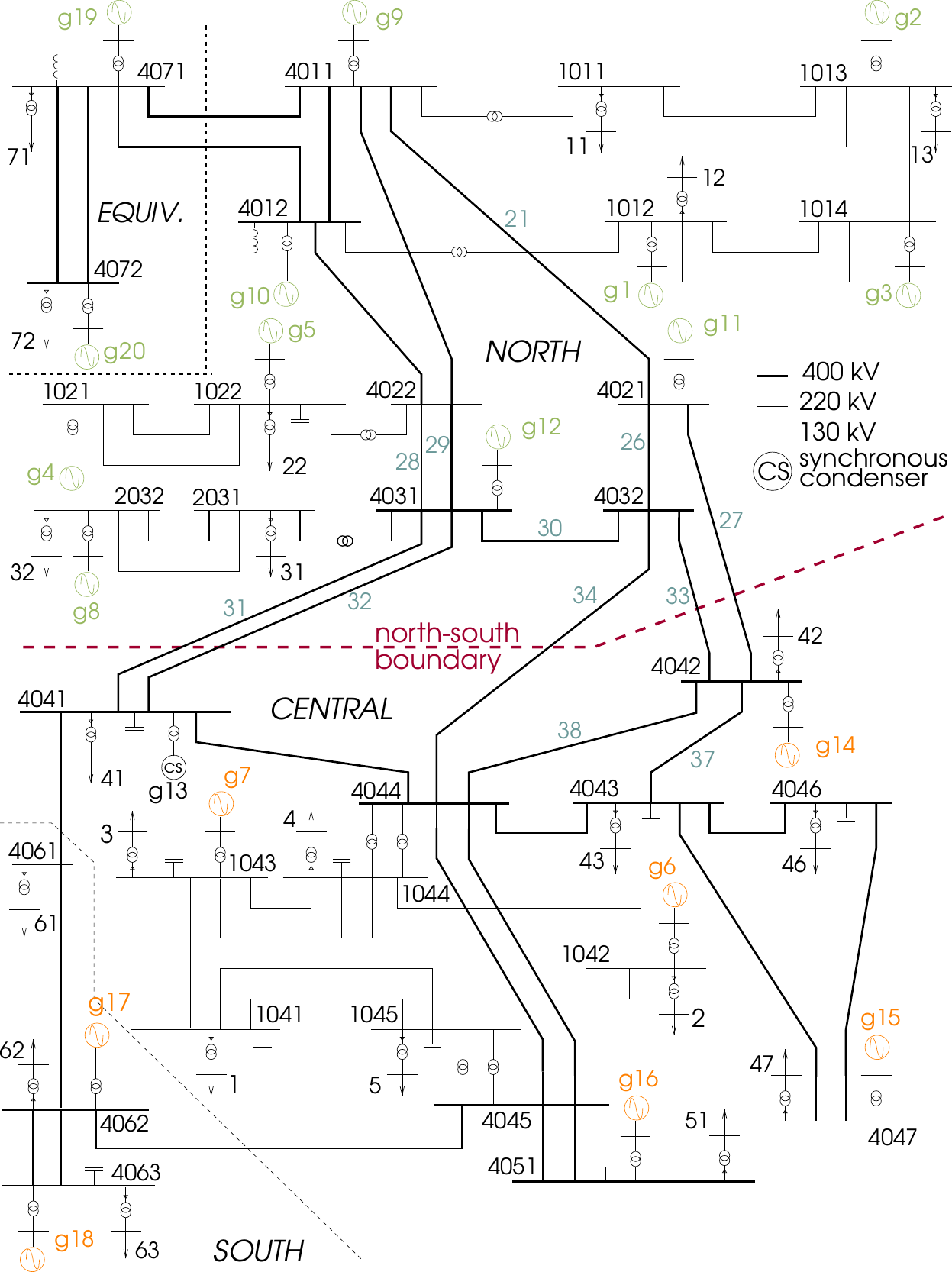}
	\caption[74 bus Nordic system diagram.]{74 bus Nordic system diagram. The four areas, South, Central, North, and Equiv., are labeled, and the interface between the North and Central areas is drawn in red. The generators in the North and Equiv. areas are drawn in green and the generators in the South and Central areas in orange. This figure is modified from \cite{van2015test}.}
	\label{fig:nordic}
\end{figure}

The Nordic System is a test system commonly used for voltage stability analysis and security assessment \cite{van2015test}. Fig. \ref{fig:nordic} shows a single line diagram of the system.
It consists of four areas, labeled South, Central, North, and Equiv. Of these, South and Central have high loads and expensive generation, while North and Equiv. have inexpensive generation. Stability constraints arise on the five long transmission lines connecting the North to the Central and South. The system has 74 buses, 102 branches, and 20 generators, and a standard frequency of 50 Hz.
We use the cost function $c_g=10p_g$ for all generators in the North and Equiv and $c_g=100p_g$ for all generators in the Central and South areas.

We consider operations during a prolonged outage of branch 34, which crosses the interface between the North and Central areas. With this line removed, the original system configuration does not admit a feasible solution (even though a feasible solution is achieved with LFAC transmission upgrades). Therefore, to enable a quantitative comparisons, we increase the active power limits of the South and Central generators by 10\% to ensure feasible operation.

\section{Evaluating the Benefits of LFAC}\label{sec:lfac_benefits}
As mentioned in Section \ref{sec:lfac}, there are multiple benefits of LFAC, some of which are unique to the LFAC system (e.g. benefits associated with lowering the frequency), while others could be achieved by similar technology such as FACTS or HVDC (e.g. power flow control). %
We therefore assess these benefits separately and in combination. We further %
make an overall comparison with equivalent HVDC upgrades.
\subsection{Comparing Frequency Reduction and Power Flow Control} \label{sec:pq_v_f}
The frequency and power flow control, separately and in combination, can be qualitatively described as follows.
\subsubsection{Control of Power Flow and Frequency (LFAC-OPF)}%
The full control capabilities of the LFAC line modeled above includes the choice of frequency $\omega$ in the subnetwork, the active power $p_{ije}^I$ sent through the interface and the reactive power $q_{ije}^I$ injected on each side of the interface. By choosing an optimal combination of those variables, the variable frequency OPF \eqref{eq:opf} provides a solution that can be used to quantify the overall steady state system-level benefits of LFAC upgrades.
\subsubsection{Control of Power Flow (PQ-OPF)}%
To demonstrate the benefits of controlling of power flow alone, we maintain the active and reactive power injections $p_{im}^I,~p_{jm}^I,~q_{im}^I,~q_{jm}^I~\forall m\in\mathcal{I}$ of the converters as variables but fix the LFAC frequency at 50 Hz by setting $\underline{\omega_l}=\overline{\omega_l}=50$ Hz in \eqref{eq:var_f}.
The solution obtained by solving the variable frequency OPF with this additional restriction quantifies the benefits that arise from power flow control without a change in frequency.
\subsubsection{Control of Frequency (F-OPF)}%
To demonstrate the benefits of frequency control in absence of direct power flow control, we add three constraints,
\begin{align}
    &q_{im}^I = q_{jm}^I &\forall (i,j)\in\mathcal{N}_m^I,m\in\mathcal{I} \label{eq:q_bal}\\
    &V_{i} = V_{j} &\forall (i,j)\in\mathcal{N}_m^I, m\in\mathcal{I} \label{eq:v_mag}\\
    &\theta_{i} = \theta_{j} &\forall (i,j)\in\mathcal{N}_m^I, m\in\mathcal{I} \label{eq:v_ang}
\end{align}
Here, \eqref{eq:q_bal} enforces the balance of reactive power at each converter, while \eqref{eq:v_mag}, \eqref{eq:v_ang} force the voltage angle and magnitudes to coincide on both sides of the converters. These constraints remove the ability to directly control active and reactive power and instead determine the flows implicitly through the AC power flow equations. In this case, the LFAC branches behave as normal AC branches with variable (frequency-dependent) line parameters, and the solution to the variable frequency OPF quantifies the benefits of changing the frequency alone, without the benefits of power flow control. %

\begin{figure}[!t]
	\centering
	\def\svgwidth{\textwidth}
	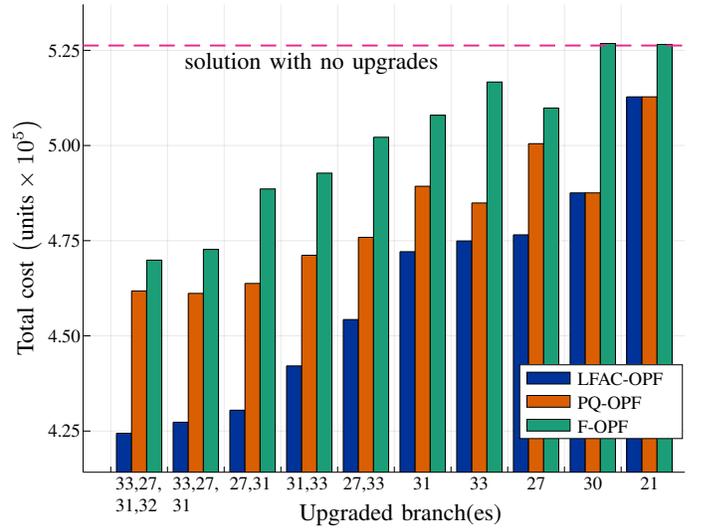
	\caption{Minimum generation cost with point-to-point LFAC upgrades. The x-axis represents the upgraded branches in each upgrade scenario. Branches 27, 31, 32 and 33 are boundary lines.
	For each upgrade, the blue, orange and green bars show the minimium cost for the LFAC-OPF, PQ-OPF and F-OPF, respectively.
	The dashed line shows the minimum cost with no upgrades.
	}
	\label{fig:nordic_srch}
\end{figure}

To compare the three different control options LFAC-OPF, PQ-OPF and F-OPF, we consider point-to-point overhead line upgrades for each of the lines across the North-South boundary and a selected set of upgrades with two, three, and four point-to-point LFAC overhead lines. For each configuration, we solve the variable frequency OPF with each of the three types of control capability. %
Fig. \ref{fig:nordic_srch} shows the minimum cost for each configuration and control capability.
We first observe that both the PQ-OPF (with power flow control) and the F-OPF (with frequency control) reduce the amount of generation in the South-Central area for all upgrades that involve boundary lines. However, the LFAC-OPF (with combined power flow and frequency control) significantly outperforms the solutions from both the PQ-OPF and F-OPF for all boundary line upgrades. This indicates that both power flow control and frequency reduction are important for the boundary lines, which have active angle difference constraints at 50 Hz. In particular, the difference between the PQ-OPF and LFAC-OPF represents the improvement that is only achieved by lowering the frequency and mitigating the angle difference constraints.

For the LFAC configurations that do not involve any boundary lines (upgrade of lines 30 and 21) we observe that the F-OPF does not reduce cost relative to the original system (without upgrades). This indicates that the benefit of frequency control on those lines, which are not stability constrained, is negligible. However, the PQ-OPF and LFAC-OPF still reduce cost relative to the original system solution, showing that LFAC upgrades provide benefits through power flow control.

In addition to the above results, we report the optimal frequencies for each LFAC line in each upgrade configuration in Table \ref{tab:opf_f}. We include the optimal frequencies both for the LFAC-OPF and the F-OPF.
The LFAC-OPF solutions for boundary line upgrades have optimal frequencies between 19 Hz and 36 Hz, showing that lowering the frequency is advantageous to allow increased power flow. For non-boundary line 30, which is not angle-constrained, the optimal choice of frequency is to remain at the standard 50 Hz.
Interestingly, the F-OPF solutions mostly show lower optimal frequencies than the LFAC-OPF solutions, with the optimal frequencies of most boundary lines ranging from 1.60 to 19.65 Hz. This indicates that in absence of power flow control, it is beneficial to lower the frequency more to encourage higher flow on the LFAC line. We also observe a higher spread in frequencies for the case with four LFAC lines, where one of the lines remains at 50 Hz, which again indicates that frequency control is used to implicitly control power flow.

\begin{table}[!t]
\caption{Optimal frequencies of the point-to-point LFAC upgrades in the LFAC-OPF and F-OPF solutions for the Nordic system example shown in Fig. \ref{fig:nordic_srch}.}
\label{tab:opf_f}
\footnotesize
\begin{tabular}{@{}lll@{}}
\toprule
\multirow{2}{*}{\shortstack[l]{upgraded\\branch(es)}} & \multicolumn{2}{l}{optimal frequency of each branch (Hz)} \\\cmidrule(lr){2-3}
										& LFAC-OPF											& F-OPF \\\midrule
33, 27, 31, 32      & 35.64, 28.61, 19.81, 19.81 		& 1.600, 1.824, 50.00, 4.940 \\
33, 27, 31          & 30.85, 28.10, 20.42 					& 1.583, 1.823, 4.852\\
27, 31              & 25.10, 20.49 									& 9.918, 12.41 \\
31, 33              & 21.23, 27.89 									& 11.59, 13.84 \\
27, 33              & 22.59, 31.31 									& 1.733, 5.469 \\
31                  & 21.28 												& 16.35 \\
33                  & 28.00 												& 19.65 \\
27                  & 20.85 												& 11.89 \\
30                  & 50.00 												& 50.00 \\
21                  & 34.54  												& 40.85 \\ \bottomrule
\end{tabular}
\end{table}

\subsection{Comparing LFAC and HVDC}
Since HVDC is a widely used, mature technology for point-to-point transmission, we evaluate the proposed LFAC transmission against a comparable HVDC upgrade. Here, we first discuss the details of the HVDC modeling and how the line limits may change when we convert an existing AC corridor from AC to HVDC. We then present simulation results for the same upgrades discussed in the previous subsection. %

The variable frequency formulation holds as frequency decreases towards the lower limit of 0 Hz, but at the limit, several features change. The voltage angle loses its significance, making the reactive power zero and the active power dependent only on the voltage magnitude. The voltage and current waveforms become constant values equal to the AC rms values.
For HVDC lines, we thus replace (\ref{eq:p_o},\ref{eq:q_o}) by the following power flow equation and remove the variable $q^E_{ije}$,
\begin{align}
    p^E_{ije}\!&\!=\!\frac{k_\mathrm{cond}\!\cdot\! k_\mathrm{ins}^2}{\sqrt{3}}\Bigg(\!V_i^2\left(\!G_{ije}\!+\!\frac{4}{3}G_{ije}^\mathrm{sh}\!\right)\!-\!V_iV_jG_{ije}\!\Bigg),\nonumber\\
    &\specialcell{\hfill\forall ije\in\mathcal{E}_l, l\in\mathcal{S}}\label{eq:dcflow_f}.
\end{align}

\begin{figure}[!t]
	\centering
	\def\svgwidth{\textwidth}
	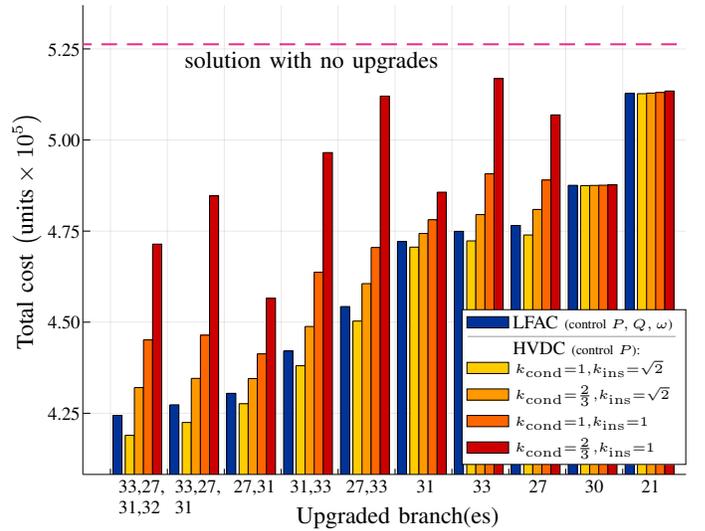
	\caption{Minimum generation cost resulting from various point-to-point LFAC and HVDC upgrades. %
	The blue bars show the results for the LFAC-OPF, which correspond to the blue bars in Fig. \ref{fig:nordic_srch}. The other bars show the results for HVDC configurations with four different combinations of the parameters $k_\mathrm{cond}$ and $k_\mathrm{ins}$.
	The dashed line shows the minimum cost with no upgrades.}
	\label{fig:nordic_hvdc}
\end{figure}

In this equation, the two parameters $k_\mathrm{cond},~k_\mathrm{ins}$ reflect two features that are relevant when comparing AC and DC power flow on equivalent physical infrastructure.
First, repurposing an existing AC line for DC transmission may result in a change in conductor utilization. A single circuit AC branch with three conductors would typically use only two conductors when operated as a bipolar DC circuit, leading to a conductor utilization of $k_\mathrm{cond}=2/3$ \cite{LARRUSKAIN20111341,Reed2019}. However, if special schemes are used \cite{Barthold2005,Xu2014} or if the upgrade involves a double circuit branch with six conductors which can use all six in three bipolar DC circuits, then $k_\mathrm{cond}=1$.
Second, for an AC circuit, insulation is designed for the peak voltage, which is larger than the rms by a factor of $\sqrt{2}$. Hence, the DC transmission voltage can be up to $\sqrt{2}$ times larger than the AC rms for the same insulation and tower geometry. This is limited, however, by DC surface and surge insulation requirements as well as environmental conditions \cite{Hausler1997,Clerici1991,George2018,IEC60815-4,LARRUSKAIN20111341}. We consider two cases. If the DC voltage remains the same as the AC voltage, we set $k_\mathrm{ins}=1$. If the DC voltage increases to the peak voltage, we set $k_\mathrm{ins}=\sqrt{2}$. We neglect the difference of resistance due to the skin effect because the difference has been shown analytically to be less than several percent for overhead lines \cite{Ngo2016PowSys}, a result confirmed by empirical data in conductor data sheets \cite{generalcable2017}.

To compare LFAC and HVDC upgrades, we consider each point-to-point upgrade with the combinations of $k_\mathrm{ins}$ and $k_\mathrm{cond}$ discussed above. We plot the results in Fig. \ref{fig:nordic_hvdc}, alongside the results obtained for the LFAC-OPF with full control of frequency, active and reactive power. We observe that for every upgrade, the optimal objective values with LFAC are comparable to the highest-capacity HVDC configuration ($k_\mathrm{ins}=\sqrt{2}, k_\mathrm{cond}=1$) and lower than each of the other HVDC configurations. These results demonstrate that LFAC can closely match---and often exceed---the power flow benefits of HVDC while bringing advantages such as well-established protection devices. We note that on the non-boundary lines 21 and 30 which are not loaded to their limits, the reduction in cost arises from increased ability to control power flow, rather than from increases in the transmission capacity of the upgraded lines. As a result, the configuration makes negligible difference for these non-congested lines.

\subsection{Effects of Problem Complexity on Solver Performance}
Because of the additional complexity of this problem compared with the standard AC OPF, we investigate the results in terms of solution times, local solutions, and failure to find feasible solutions. First, it is important to note that the added complexity scales with the size and number of variable frequency subnetworks and converters, not with the size of the overall network. Multi-frequency systems in practice involve only several lines with frequency as an optimization variable compared to a large fixed-frequency network. For branches and buses with standard frequencies, $\omega$ is a constant, and the corresponding constraints represent standard AC power flow.

The solver performance on the various problems introduced here is shown in Table \ref{tab:solve_time}. It is evident that the addition of both power flow control and frequency as variables increase the complexity of the OPF problem, leading to more iterations and longer solve times. However, the solution times are not prohibitive, as all problems solve within less than 11 s.

\begin{table}[!t]
	\caption{Solve time and number of iterations across all the point-to-point LFAC and HVDC and multi-terminal LFAC upgrades shown in Fig. \ref{fig:nordic_srch}, \ref{fig:nordic_hvdc} and \ref{fig:nordic_rel}, from five trials on each configuration.}
	\label{tab:solve_time}
	\footnotesize
\begin{tabular}{@{}llllllll@{}}
\toprule
\multicolumn{2}{l}{\multirow{2}{*}{type of model}} & \multicolumn{3}{c}{solver CPU time (s)} & \multicolumn{3}{c}{iterations} \\ \cmidrule(l){3-5}\cmidrule(l){6-8}
\multicolumn{2}{l}{}                                 & mean     & max      & std dev  & mean     & max    & std dev    \\ \midrule
\multicolumn{2}{l}{\shortstack[l]{standard\\AC OPF}}           & 0.5025   & 0.6220   & 0.05654  & 61       & 61     & 0          \\\midrule
\multicolumn{8}{l}{\scriptsize AC OPF with point-to-point upgrades:}\\
    & LFAC-OPF      & 2.956    & 10.79    & 2.562    & 244.3    & 787    & 204.7      \\
                                     & PQ-OPF        & 1.395    & 3.499    & 0.7686   & 129.2    & 292    & 69.26      \\
                                     & F-OPF         & 1.804    & 4.154    & 0.8709   & 151.6    & 272    & 56.66      \\
																		 & HVDC          & 0.868    & 2.135    & 0.3195   & 88.36    & 175    & 31.94      \\\midrule
\multicolumn{8}{l}{\scriptsize AC OPF with multi-terminal upgrades:}\\
& LFAC-OPF    & 2.655    & 5.496    & 1.165    & 253.4    & 398    & 92.21      \\ \bottomrule
\end{tabular}
\end{table}

Each of the LFAC-OPF solutions in Section \ref{sec:lfac_benefits} is analyzed for local solutions by solving the OPF with allowable LFAC frequencies in smaller (5 Hz) ranges throughout the full 0-50 Hz range. In all cases, the solution obtained when the frequency is allowed its full range matches the minimum of the solutions under smaller ranges. In addition, choosing different starting points for the LFAC frequency in increments of 10 Hz from 0 to 50 Hz results in identical solutions in all LFAC-OPF cases. From this analysis, we have no indication that additional complexity causes the solver to converge to suboptimal local minima in the point-to-point LFAC upgrades presented here.

\begin{figure}[!t]
	\centering
	\def\svgwidth{\textwidth}
	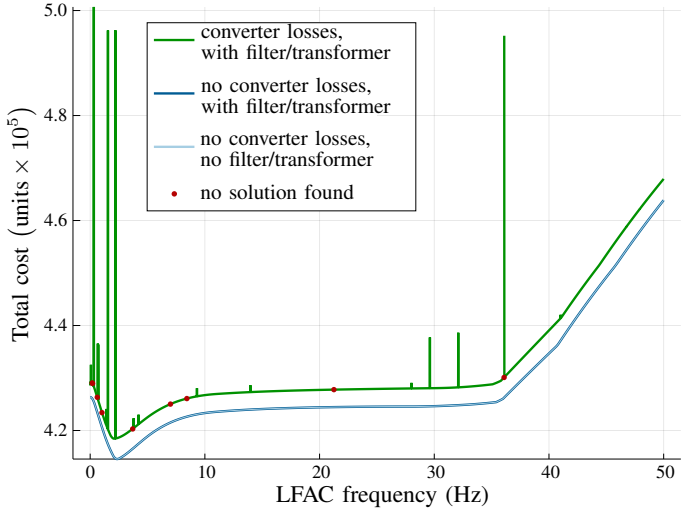
	\caption{Minimum generation cost OPF solution as frequency varies in a multiterminal LFAC upgrade. The green line shows the solution value when the full converter loss model and filter or transformer branches are included, while the dark and light blue represent the solution neglecting converter losses, with and without the filter/transformer, respectively.}
	\label{fig:loss_comparison}
\end{figure}

The converter loss model is a significant source of model complexity, particularly \eqref{eq:a_mabs}. To analyze the effects of this complexity, we consider a multi-terminal LFAC upgrade involving six converters, corresponding to configuration 10 in Table \ref{tab:mt_config}. We perform what we will refer to as a \emph{frequency sweep}: For each frequency between 0 and 50 Hz in increments of 0.01 Hz, we fix the LFAC frequency as a parameter and solve three different versions of the LFAC-OPF. The results are plotted in Fig. \ref{fig:loss_comparison}, and the three different solutions correspond to a model that considers the full converter loss model and filter or transformer branches (green), a model that does not consider converter losses but includes the filter or transformer branches (dark blue), and a model without either the converter losses or filter or transformer (light blue). We first observe that the difference between the two blue solutions is negligible (they overlap in the plot), indicating that the relatively small filters or transformers of an MMC have negligible impact on the total cost. However, there is a more significant difference between the models with (green) and without (blue) converter losses, where the converter losses slightly increase the cost.
The additional complexity of including converter losses in the problem leads in several cases to local solutions or failure to find a solution due to numerical errors, as indicated by the spikes on the green curve. The number of these cases is small, with 9 unsolved and 15 apparent local minima compared to the total of 5000 frequency steps. Solving the LFAC-OPF with frequency as a variable in the range of 0 to 50 Hz finds the minimum at 2.10 and 2.35 Hz with and without converter losses, which matches the minimum cost frequencies found in the frequency sweep above. This demonstrates that the LFAC-OPF  does not converge to a suboptimal local minimum in this case.

\section{Selecting Optimal Transmission Frequencies}\label{sec:optimal_frequency}
We next analyze how the maximum transmission capacity changes with frequency and how the optimal transmission frequency changes with topology and operating conditions.

\subsection{Maximum Transmission Capacity for Different Frequencies}
\label{sec:f_dep}
\begin{figure}[!t]
	\centering
	\def\svgwidth{\textwidth}
	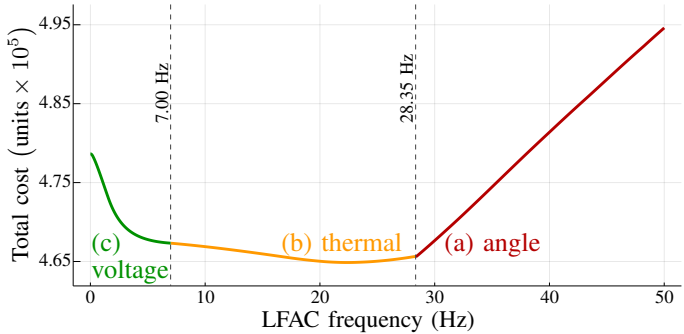
	\caption{Minimum objective value for a multi-terminal LFAC system (configuration 4 in Table \ref{tab:mt_config}) as frequency varies. Converter losses are neglected.}
	\label{fig:nordic_mt_sweep}
\end{figure}

We first analyze the objective values and limiting constraints at different frequencies.
For this analysis, we define a multi-terminal variable frequency subnetwork encompassing branches 21, 27, and 37, with converters interfacing the subnetworks at buses 4011, 4021, and 4043. This represents an upgrade of existing lines into a long LFAC corridor with several connections to the 50 Hz network along its length.
Because the aim of this analysis is to observe how the solution changes with frequency and the active power losses in the converters do not vary significantly with frequency, as shown in Fig. \ref{fig:loss_comparison}, the losses are neglected here by setting $p_{im}^{\textrm{cond}}=p_{jm}^{\textrm{cond}}=p_{im}^{\textrm{sw}}=p_{jm}^{\textrm{sw}}=0$ in \eqref{eq:interface_pbal}.
We fix the frequency, solve the OPF, and then repeat for a range of frequencies from 50 to 0 Hz at intervals of 0.01 Hz. Fig. \ref{fig:nordic_mt_sweep} shows the resulting objective value as a function of the frequency.

At an LFAC frequency of 50 Hz, we see a reduction in cost of almost 6\% relative to the solution with no upgrades. This vertical offset represents the improvement that is achieved from active and reactive power control alone, without lowering the frequency. Next, as the frequency is lowered from 50 Hz, three ranges of dominant active constraints are evident.
\begin{enumerate}[label=(\alph*)]
	\item \emph{Angle constrained:} From 50 Hz down to 28.35 Hz, the system is limited by one or more angle constraints on lines 21, 27, and non-LFAC line 33. At lower frequencies, the angle constraints become less restrictive than the thermal limits.
	\item \emph{Thermally constrained:} From 28.44 Hz to 1.57 Hz, power flow on lines 21 and 27 is at the thermal limits.
    While the lines are at their thermal limits, changes in frequency have only small, indirect effects through the reactive power, making this region mainly flat. The optimal frequency lies within this range.
	\item \emph{Voltage drop constrained:} At 7.00 Hz, the voltage at the LFAC bus south of the boundary reaches its lower voltage limit of 0.9 p.u., with the northern LFAC buses at their 1.1 p.u. upper voltage limits.
    This continues at lower frequencies, becoming most severe approaching 0 Hz, where power flow is determined only by voltage magnitudes and resistance.
\end{enumerate}
This example, typical of a potential multi-terminal LFAC upgrade, illustrates a pattern of frequency dependence which closely matches the analytical results for a single line in \cite{lfacmodeling}. %

\subsection{Optimal Frequency for Various Topologies}\label{sec:multiterminal_topologies}
\begin{figure}[!t]
	\centering
	\def\svgwidth{0.95\textwidth}
	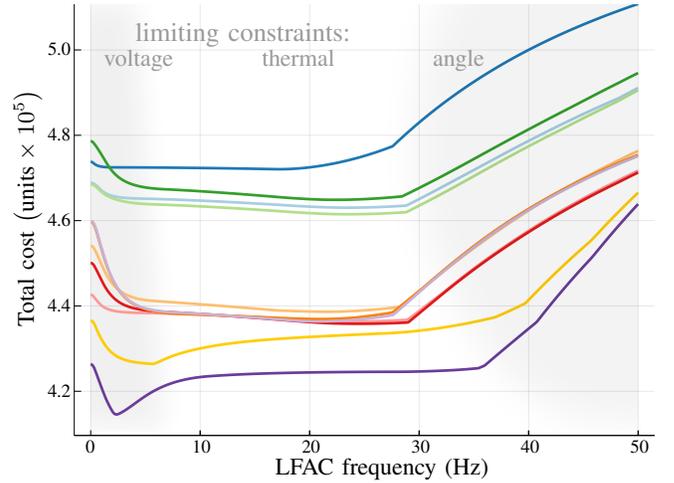
	\caption{Optimal objective value versus frequency for the multi-terminal LFAC upgrades in Table \ref{tab:mt_config}, neglecting converter losses.
    }
	\label{fig:nordic_rel}
\end{figure}

To investigate how the maximum transmission capacity and optimal frequency change with topology, we consider 11 LFAC reconfigurations of the Nordic system. Maintaining all lines and buses in place, we add converters to create multi-terminal LFAC networks. Each upgrade involves between two and eight lines (at least one or more across the inter-area boundary) and between two and six converters. The configurations are summarized in Table \ref{tab:mt_config} with further details listed in the Appendix. We solve the variable frequency OPF and give the minimum cost and minimizing frequencies in Table \ref{tab:mt_config}. We also solve the OPF over a sweep of frequencies for each configuration and plot the results in Fig. \ref{fig:nordic_rel}.

\begin{table}[!t]
\caption{Properties and results for several configurations of the Nordic system with a multi-terminal LFAC subnetwork.}
\small
\begin{tabular}{rrrrrrrr}
\toprule
\rot{\scriptsize color} & \rot{\scriptsize configuration} & \rot{\scriptsize \# lines} & \rot{\scriptsize \# N-S lines} & \rot{\scriptsize \# converters} & \rot{\scriptsize \shortstack[l]{minimum\\cost\\($10^5$ units)}}& \rot{\scriptsize \shortstack[l]{\% improved\\from upgrade}} & \rot{\scriptsize \shortstack[l]{optimal\\frequency\\(Hz)}}\\ \midrule
\textcolor{color01}{\rule[0.5mm]{3mm}{0.5mm}}& 1  & 2 & 1 & 3 & 4.6304 & 12.01\% & 23.51 \\
\textcolor{color02}{\rule[0.5mm]{3mm}{0.5mm}}& 2  & 2 & 1 & 2 & 4.7205 & 10.30\% & 17.22 \\
\textcolor{color03}{\rule[0.5mm]{3mm}{0.5mm}}& 3  & 3 & 1 & 4 & 4.6149 & 12.31\% & 23.51 \\
\textcolor{color04}{\rule[0.5mm]{3mm}{0.5mm}}& 4  & 3 & 1 & 3 & 4.6488 & 11.66\% & 22.37 \\
\textcolor{color05}{\rule[0.5mm]{3mm}{0.5mm}}& 5  & 3 & 1 & 4 & 4.3634 & 17.09\% & 23.89 \\
\textcolor{color06}{\rule[0.5mm]{3mm}{0.5mm}}& 6  & 4 & 1 & 5 & 4.3582 & 17.19\% & 24.46 \\
\textcolor{color07}{\rule[0.5mm]{3mm}{0.5mm}}& 7  & 4 & 1 & 4 & 4.3863 & 16.65\% & 21.51 \\
\textcolor{color08}{\rule[0.5mm]{3mm}{0.5mm}}& 8  & 3 & 1 & 6 & 4.3692 & 16.98\% & 20.93 \\
\textcolor{color09}{\rule[0.5mm]{3mm}{0.5mm}}& 9  & 4 & 1 & 6 & 4.3638 & 17.08\% & 20.85 \\
\textcolor{color10}{\rule[0.5mm]{3mm}{0.5mm}}& 10 & 8 & 3 & 6 & 4.1456 & 21.23\% & 2.351 \\
\textcolor{color11}{\rule[0.5mm]{3mm}{0.5mm}}& 11 & 8 & 3 & 5 & 4.2641 & 18.97\% & 5.662 \\ \bottomrule
\end{tabular}%
\label{tab:mt_config}
\end{table}

From Fig. \ref{fig:nordic_rel}, it is evident that the configurations involving one boundary line have frequency dependence similar to that described in Section \ref{sec:f_dep}, which corresponds to configuration 4. %
However, each configuration has a different vertical offset at 50 Hz, showing different gains from power flow control. The optimal frequencies are similar but not identical, and the effects of the voltage drop limitations at low frequencies vary in severity across these configurations. These differences illustrate the effects of different layouts of lines and placement of converters.
We also note that the largest LFAC configurations 10 and 11 show a distinct shape. %
In the thermally constrained region, only some of the lines are at their thermal limits, and the improvement is only 80-85\% of the full value.
Due to more complex interactions involving voltage constraints, when the frequency is lowered further, more power flows on line 30 up to the thermal limit, and the optimal frequency is found.

\subsection{Frequency Dependence With Different Operating Conditions}%
The various configurations in Table \ref{tab:mt_config} have different optimal frequencies. This raises the question of whether a LFAC system should be designed with one fixed frequency or whether the frequency should be allowed to change in response to changing system conditions, such as every day or hour. We provide an example of how changing the frequency can provide system benefits.

\begin{figure}[!t]
	\centering
	\scriptsize
	\def\svgwidth{\textwidth}
	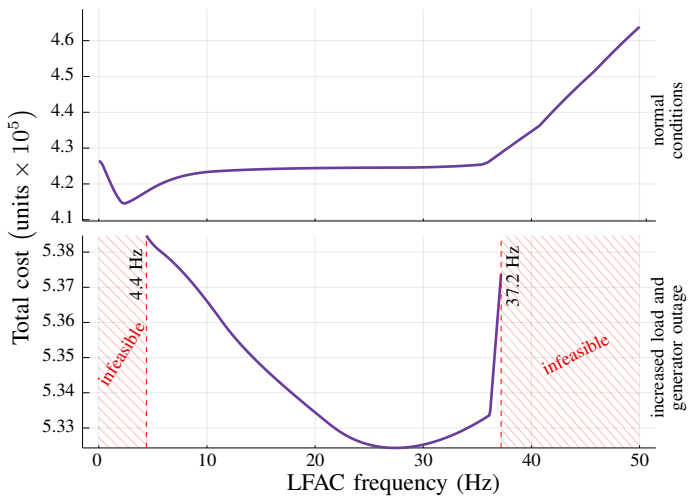
	\caption{Minimum generation cost in the South-Central region versus frequency for multi-terminal LFAC configuration 10. Top: Original operating conditions (corresponding to results in Fig. \ref{fig:nordic_rel}). Bottom: Heavily loaded operating conditions with a generator 12 outage and a 13\% increase in load. Converter losses are neglected.}
	\label{fig:nordic_g12out}
\end{figure}

For our analysis, we consider configuration 10, which has the lowest cost and the lowest optimal frequency in Table \ref{tab:mt_config}. We analyze how the optimal frequency changes for a different operating condition, when the system is heavily stressed. To this end, we increase all demand by 13\% and assume an outage of generator 12 in the North. We then solve the OPF for a sweep of frequencies from 0 to 50 Hz. Fig. \ref{fig:nordic_g12out} compares the solutions of the frequency sweeps in the original (top) and heavily loaded (bottom) operating conditions. %
In contrast to the original operating conditions with optimal frequency at 2.35 Hz, the heavily loaded operating condition admits no feasible solution below 4.4 Hz or above 37.2 Hz. If the system were designed based on the optimal frequency of 2.35 Hz at the first operating conditions, it would not have a feasible solution in these conditions.
These results demonstrate that power flow can change significantly with frequency. In some cases only certain frequencies lead to feasible solutions, and changes in operating conditions can significantly change the optimal frequency or the range of feasible frequencies. It is advantageous to take advantage of the flexibility that an LFAC system gives to achieve feasible and optimal solutions under varying load conditions and contingencies.

\section{Conclusions}\label{sec:conclusions}
In this paper, we present a variable frequency optimal power flow formulation based on AC power flow equations in polar form and a converter model.
We discuss its implementation and results for a realistic stability-constrained system with active angle difference constraints.
We conclude that the system can benefit from the power flow control that LFAC converters allow, and the increased power transfer capacity from low frequencies on stability-constrained lines dramatically improves operation of the whole system. %
Furthermore, we find that the LFAC solutions compare favourably to the results achieved with comparable HVDC upgrades.
We further analyze the relationship between maximum power transfer and choice of frequency, %
demonstrate changes in this dependence across system topologies, and show that a frequency which is optimal at one operating point may be far from optimal and not even admit a feasible solution at another.

The above results underscore that LFAC could become a promising technology for transmission system upgrades. Furthermore, we observe that models and tools such as those presented here are essential for analyzing and understanding the design and operation of LFAC systems. %
Future work includes the consideration of LFAC in N-1 contingency analysis and a more thorough investigation of different operating conditions, including systems with high penetrations of renewable energy. Furthermore, we are interested in understanding how LFAC technology can enable undergrounding of the transmission system (and in particular long underground cables) in regions with exposure to extreme weather and wildfire risk. Additional research directions would include long-term cost-benefit analysis for investment decisions, analysis of different converter control schemes and more extensive analysis of power system dynamic performance with embedded LFAC systems.

\appendix

The multi-terminal LFAC upgrade configurations in Table \ref{tab:mt_config} are described in more detail in Table \ref{tab:mt_details}. These configurations rely only on existing lines and buses, shown in the table. All buses at which converters are added to create an interface between the 50 Hz and LFAC system are given in the third column. The final column lists other buses which are either fully converted to LFAC or split to accommodate 50 Hz and LFAC without a connection between the two (labeled ``split").

\begin{table}[!t]
\caption{Multi-terminal LFAC upgrades for the Nordic system considered in Section \ref{sec:multiterminal_topologies}. Each configuration uses existing lines and buses, adding only frequency converters to create an LFAC subnetwork. In some cases, buses are split between the 50 Hz and LFAC subnetworks.}\label{tab:mt_details}
\scriptsize
\begin{tabular}{@{}Q{1cm}L{1.5cm}L{2.25cm}L{2.25cm}@{}}
\toprule
configuration & LFAC branches                  & buses with converters             & other buses converted to LFAC   \\\midrule
1             & 21, 27                         & 4011, 4021, 4042                  &                                 \\\cline{1-4}
2             & 21, 27                         & 4011, 4042                        & 4021 (split)                    \\\cline{1-4}
3             & 21, 27, 37                     & 4011, 4021, 4042, 4043            &                                 \\\cline{1-4}
4             & 21, 27, 37                     & 4011, 4021, 4043                  & 4042 (split)                    \\\cline{1-4}
5             & 21, 26, 27                     & 4011, 4032, 4042, g11             & 4021                            \\\cline{1-4}
6             & 21, 26, 27, 37                 & 4011, 4032, 4042, 4043, g11       & 4021                            \\\cline{1-4}
7             & 21, 26, 27, 37                 & 4011, 4032, 4043, g11             & 4021, \hbox{4042 (split)}       \\\cline{1-4}
8             & 21, 26, 27                     & 4011, 4032, 4042, 42, g11, g14    & 4021              							 \\\cline{1-4}
9             & 21, 26, 27, 37                 & 4011, 4032, 4043, 42, g11, g14    & 4021, \hbox{4042 (split)}			 \\\cline{1-4}
10            & 21, 26, 27, 30, 33, 34, 37, 38 & 4011, 4031, 4042, 4043, 4044, g11 & 4021, 4032                      \\\cline{1-4}
11            & 21, 26, 27, 30, 33, 34, 37, 38 & 4011, 4031, 4042, 4043, 4044      & 4021, 4032                      \\\bottomrule
\end{tabular}%
\end{table}

\section*{Acknowledgment}
The authors would like to thank our project partners at New York Power Authority (NYPA) and in particular Shayan Behzadirafi and  Greg Pedrick, as well as Giri Venkataramanan at University of Wisconsin--Madison for the discussions that helped develop and improve this work. The authors would also like to thank the members of the Wisconsin Electric Machines and Power Electronics Consortium (WEMPEC).

\ifCLASSOPTIONcaptionsoff
  \newpage
\fi

\bibliographystyle{IEEEtran}
\bibliography{IEEEabrv,refs}

\begin{IEEEbiography}[{\includegraphics[width=1in,height=1.25in,clip,keepaspectratio]{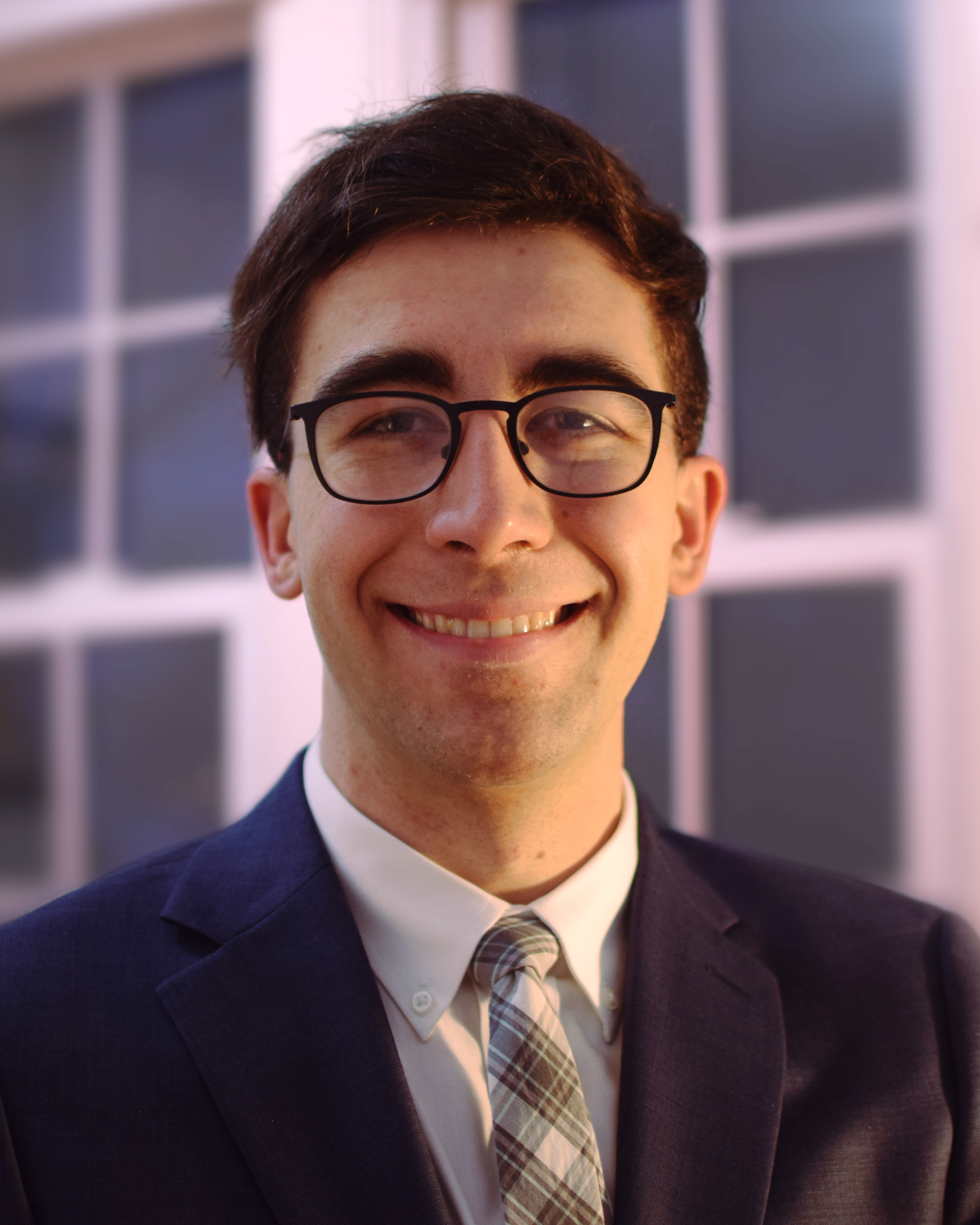}}]{David Sehloff}
received B.S. degrees in electrical engineering and in systems science and engineering from Washington University in St. Louis in 2016 and the M.S. degree in electrical engineering in 2018 from the University of Wisconsin-Madison, where he is currently pursuing the Ph.D. degree in electrical engineering. He is a student member of the Wisconsin Electric Machines and Power Electronics Consortium (WEMPEC). His research interests include modeling and optimization for power systems.
\end{IEEEbiography}

\begin{IEEEbiography}[
{\includegraphics[width=1in,height=1.25in,clip,keepaspectratio]{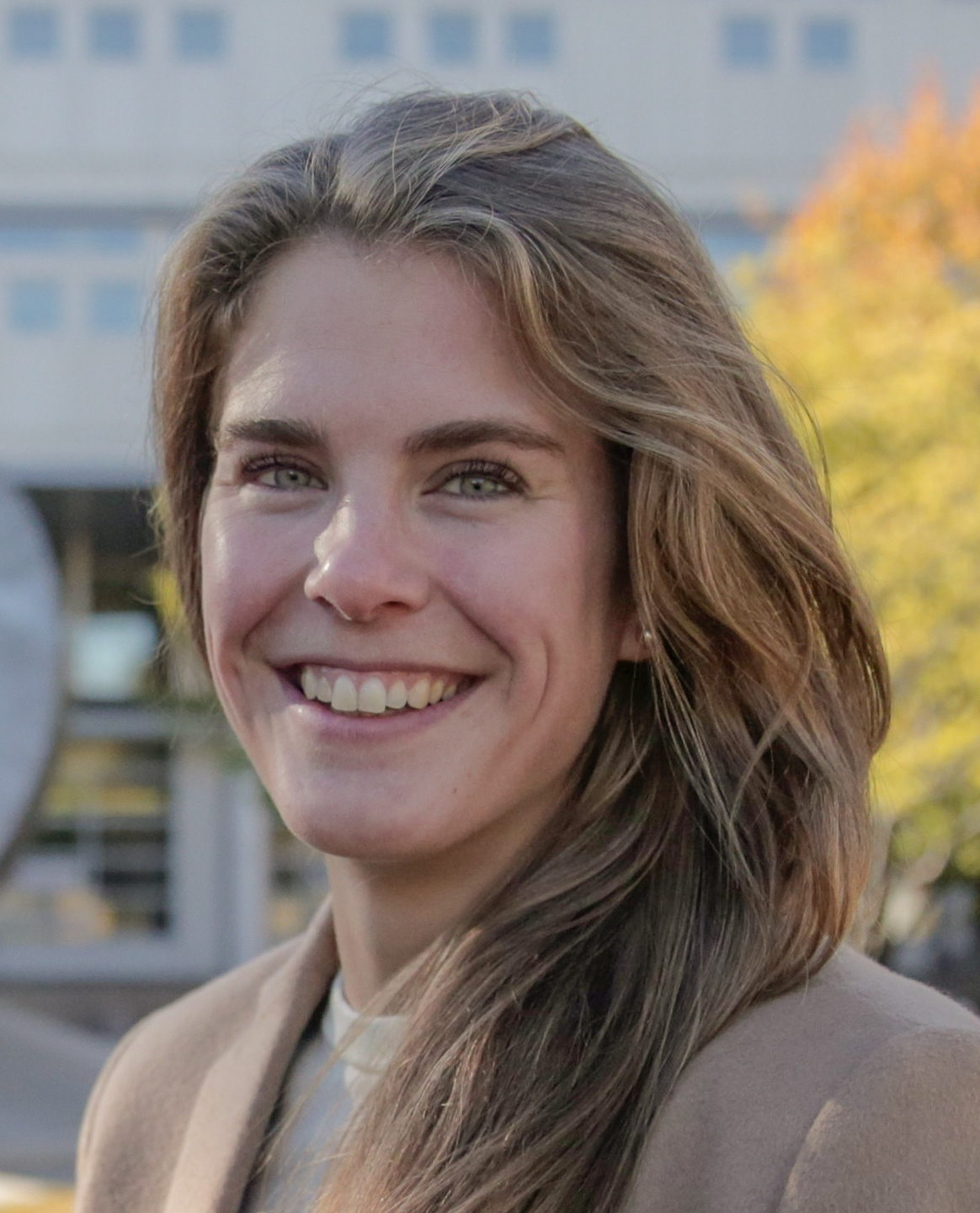}}]{Line Roald}  is an Assistant Professor and Grainger Institute Fellow in the Department of Electrical and Computer Engineering in University of Wisconsin--Madison.
She received her Ph.D. degree in Electrical Engineering (2016) and M.Sc. and B.Sc. degrees in Mechanical Engineering from ETH Zurich, Switzerland. She is the recipient of an NSF CAREER award, and the UW Madison ECE Outstanding Mentoring Award. Her research interests focus on modeling and optimization of energy systems, with a particular focus managing uncertainty and risk from renewable energy variability and large-scale outages.
\end{IEEEbiography}

\end{document}